\documentclass[preprintnumbers,amsmath,amssymb,12pt,floatfix,epsfig,showpacs,showkeys,A4]{revtex4}
\usepackage{dcolumn}
\usepackage{bm}
\usepackage{stmaryrd}
\usepackage[dvips]{color}
\usepackage{graphicx}
\usepackage{ulem}
\leftmargin 0.5in \baselineskip=24ptplus.5ptminus.2pt \vspace*{0.5in} \large
\begin{document}

\title{Coupled-channels analyses for
$^{9,11}$Li + $^{208}$Pb fusion reactions with multi-neutron transfer
couplings}

\author{Ki-Seok Choi and Myung-Ki Cheoun}
\thanks{\textrm{e-mail:} cheoun@ssu.ac.kr (Corresponding Author) }
\address{Department of Physics and OMEG Institute, Soongsil University, Seoul  156-743, Korea}

\author{ W. Y. So}
\address{Department of Radiological Science, Kangwon National University at Dogye, Samcheok 245-905, Korea}

\author{ K. Hagino}
\address{Department of Physics, Tohoku University, Sendai 980-8578, Japan}
\address{Research Center for Electron Photon Science, Tohoku University, 1-2-1
Mikamine, Sendai 982-0826, Japan}

\author{K. S. Kim}
\address{School of Liberal Arts and Science, Korea Aerospace University, Koyang 412-791, Korea}
\date{\today}

\begin{abstract}
We discuss the role of two-neutron transfer processes in the
fusion reaction of the $^{9,11}$Li + $^{208}$Pb systems.
We first analyze the $^{9}$Li + $^{208}$Pb reaction by taking into
account the coupling to the
$^{7}$Li + $^{210}$Pb channel.
To this end, we assume that two neutrons are directly transferred to a single
effective channel in $^{210}$Pb and solve the coupled-channels equations
with the two channels. By adjusting the coupling strength and the
effective $Q$-value, we successfully reproduce the experimental fusion
cross sections for this system.
We then analyze the $^{11}$Li + $^{208}$Pb reaction in a similar manner,
that is, by taking into account three effective channels with
$^{11}$Li + $^{208}$Pb, $^{9}$Li + $^{210}$Pb, and $^{7}$Li + $^{212}$Pb
partitions.
In order to take into account the halo structure of the $^{11}$Li nucleus,
we construct the potential between
$^{11}$Li and $^{208}$Pb with a double folding procedure, while we employ
a Wood-Saxon type potential with the global Aky\"uz-Winther parameters
for the other channels.
Our calculation indicates that the multiple two-neutron transfer process plays
a crucial role in the $^{11}$Li + $^{208}$Pb
fusion reaction at energies around the Coulomb barrier.
\end{abstract}


\maketitle

\section{Introduction}

Fusion is a process in nuclear reactions in which
a projectile nucleus collides with a target nucleus and then
the two nuclei are merged into a new compound nucleus.
The compound nucleus formed in fusion reaction is in general
highly excited, and it decays by
emitting gamma ray(s), neutron(s), proton(s), and alpha particle(s).
The nuclear fusion plays an important role in the energy
generation in the stellar evolution as well as
in the quest for superheavy elements production.
See Refs. \cite{BT98,DHRS98,hagino2012subbarrier,Back14,Montagnoli17}
for recent reviews.

Fusion cross sections are strongly influenced by
the Coulomb barrier, which is constructed as a sum of the repulsive
Coulomb and an attractive nuclear potentials.
While the charge numbers of the projectile and the target nuclei provide
the strength of the repulsive Coulomb potential, the
mass numbers are related to the strength for the attractive
nuclear potential.
When the incident energy is lower
than the Coulomb barrier height,
most of the flux does not pass through
the Coulomb barrier and is scattered elastically.
In this situation, the fusion takes place only by quantum tunneling.
As the incident energy increases, the fusion cross sections also
increase~\cite{so2011effect}, and eventually coincide with the classical
fusion cross sections (see e.g., the ``Wong formula''
~\cite{PhysRevLett.31.766,PhysRevC.91.044617} obtained in the
parabolic approximation to the Coulomb barrier).


Recently, various radioisotope (RI) neutron-rich beams,
such as $^{6,8}$He, $^{9,11}$Li, $^{11}$Be and $^{16,19}$C, have been produced
thanks to the remarkable advances in the radioisotope beam technology.
The fusion process of such radioisotopes has attracted lots of attention.
In fact,
a large number of experimental works have been carried out
to measure total fusion cross sections of e.g.,
$^{6}$He + $^{209}$Bi~\cite{kolata98,hassan2006},
$^{11}$Be + $^{209}$Bi~\cite{SIGNORINI2004329},
$^{11}$Li + $^{208}$Pb~\cite{PhysRevC.87.044603},
$^6$He+$^{238}$U \cite{Raabe04}, $^{6,8}$He+$^{197}$Au \cite{Lemasson09},
and $^{15}$C+$^{232}$Th \cite{Alcorta11} systems.
Many theoretical studies have also been performed
by taking into account the characteristic features
of weakly-bound neutron-rich nuclei,
such as a halo structure and a low energy threshold for breakup processes
~\cite{Canto20061,Canto20151,PhysRevC.65.024606,PhysRevC.61.037602,Ito06}.

In this paper, we discuss fusion reactions of the
$^{9,11}$Li + $^{208}$Pb systems, in which the $^{11}$Li
nucleus is
a typical example of weakly-bound halo nuclei \cite{BE91,EBH99,HS05}.
Even though the fusion of the $^9$Li nucleus is considered to provide
reference cross sections in discussing the fusion of the $^{11}$Li
nucleus, the standard coupled-channels calculations have faced difficulties
in reproducing the experimental data. For instance,
the fusion cross sections for the $^9$Li+$^{70}$Zn system
are largely underestimated at energies below the Coulomb barrier even
if the collective excitations in the colliding nuclei as well as the
ground-state-to-ground-state two-neutron transfer channel are taken into
account \cite{Loveland06}. The fusion of the $^9$Li+$^{208}$Pb system
also shows a similar difficulty \cite{PhysRevC.87.044603,PhysRevC.80.054609}.
Our fist motivation in this work is to investigate
whether the experimental data for the $^9$Li+$^{208}$Pb system
can be accounted for when one considers
the two-neutron transfer to excited states, rather than to
the ground state.
We then discuss the fusion of the $^{11}$Li+$^{208}$Pb system, which
couples to the $^9$Li+$^{208}$Pb system. Our second motivation
in this paper is to
discuss whether one can describe the fusion of $^{11}$Li+$^{208}$Pb
and $^{9}$Li+$^{208}$Pb systems in a consistent manner by taking
into account the multi two-neutron transfer process. One-neutron transfer is also possible process.
But, for Borromean nuclei like $^{11}$Li, the one-neutron transfer is known to be much less probable than the two-neutron transfer, as confirmed in the data for $^{6}$He +$^{65}$Cu system \cite{Chat08}.

The paper is organized as follows. In Sec. II, we first analyze the
fusion of the $^9$Li+$^{208}$Pb system, by taking into account the
two-neutron transfer channel. We then discuss the fusion of
$^{11}$Li+$^{208}$Pb system in Sec. III,
using the result of the calculation for the
$^9$Li+$^{208}$Pb system. We finally summarize the paper in Sec. IV.


\section{$^9$L\lowercase{i}+$^{208}$P\lowercase{b} fusion reaction}

We first analyze the $^9$Li+$^{208}$Pb fusion reaction.
For simplicity, we ignore the effect of the collective excitations
in the colliding nuclei, which is expected to be small
for this system \cite{PhysRevC.80.054609}.
Instead, we take into account the two-neutron transfer
$^{208}$Pb($^{9}$Li,$^7$Li)$^{210}$Pb channel, whose ground-state-to-ground-state
$Q$-value is $Q_{gg}$ = +3.0 MeV.
An important fact is that the transfer to the ground state
may not be the dominant process when the transfer $Q$-value is
positive \cite{Rowley2001}.
Rather, from the viewpoint of the $Q$-value matching, the transfer to
excited states would be more plausible.
In order to investigate the effect of such process,
we here solve the coupled-channels equations
\cite{hagino2012subbarrier,HAGINO1999143}
by including the transfer channel. To this end, we follow
Ref. \cite{Rowley2001} and introduce a single effective channel
for the transfer partition.
The resultant coupled-channels equations read,
\begin{equation}
\left(\begin{array}{cc}
K+V_2(r)-E & F_{  2 \rightarrow 3 }(r)  \\
F_{ 2 \rightarrow 3 }(r) & K+V_3(r)-(E+Q_{23})  \\
\end{array}\right)
\left(\begin{array}{c}
\psi_2(r)  \\
\psi_3(r)  \\
\end{array}\right)=0,
\label{eq:3}
\end {equation}
where the channels 2 and 3 denote the $^{9}$Li+$^{208}$Pb
and the $^{7}$Li+$^{210}$Pb systems, respectively.
Here, $K$ is the kinetic energy (with the centrifugal potential)
and $V_i(r)$ ($i$=1,2) is the inter-nucleus potential for each
partition.
$Q_{23}$ is the effective $Q$-value for the two-neutron transfer process,
which is determined by fitting to the experimental fusion cross sections,
while
$F_{ 2 \rightarrow 3 }(r)$ is the coupling form factor.

In the calculations presented below, we employ
the Aky\"uz-Winther (AW) potential~\cite{akyuz1981proceedings}
for the nuclear part of $V_i(r)$, whose parameters have been
globally determined. The actual values for the parameters are given in
Table 1, together with the height, the position, and the curvature
of the Coulomb barrier.
For the coupling form factor, we use a derivative form of
the Woods-Saxon potential~\cite{dasso1985macroscopic}, that is,
\begin{equation}
F_{2\rightarrow 3}(r)=F_t \,
\frac{d}{dr}\left( \frac{1}{1+\exp((r-R_{\rm coup})/a_{\rm coup})} \right).
\label{eq:2}
\end {equation}

\begin{table}[h]
\begin{ruledtabular}
\begin{center}
\caption{The depth, the radius, and the diffuseness parameters
for the nuclear Woods-Saxon potential for each channel in
the $^{9}$Li + $^{208}$Pb reaction.
These are based on the global Aky\"uz-Winther (AW)
potential~\cite{akyuz1981proceedings}.
The barrier height, $V_b$, the barrier position, $R_b$, and the
barrier curvature, $\hbar\Omega$, are also shown for each potential.
}
\label{tab:1}
\begin{tabular}{c|ccc|ccc}
Channel  & $V_0$  (MeV) & $r_0$ (fm) &   $a_0$ (fm) & $V_b$ (MeV) &
$R_b$ (fm) & $\hbar\Omega$ (MeV)
\\
\hline
$^{9}$Li + $^{208}$Pb (channel 2)  &   47.500 & 1.177  & 0.636 & 29.1
& 11.5 & 4.07 \\
$^{7}$Li +  $^{210}$Pb (channel 3) &   47.347 & 1.177  & 0.625 &
29.6 & 11.3 & 4.71  \\
\end{tabular}
\end{center}
\end{ruledtabular}
\end{table}

\begin{table}[h]
\begin{ruledtabular}
\begin{center}
\caption{The effective transfer $Q$-value and the parameters for the
coupling form factor given by Eq. (\ref{eq:2}), obtained by fitting
the results of the coupled-channels calculations to the experimental data.
Here, $r_{\rm coup}$ is defined as $R_{\rm coup}=r_{\rm coup}
(A_p^{\frac{1}{3}}+A_t^{\frac{1}{3}})$, where $A_p$ and $A_t$ are the
mass number of the projectile and the target nuclei, respectively.}
\begin{tabular}{cccc}
$Q_{23}$ (MeV) & $F_{t}$ (MeV fm) & $r_{\rm coup}$ (fm) &   $a_{\rm coup}$ (fm)
\\ \hline
$-$3.204 & 51.367   &  1.357 & 0.264               \\
\end{tabular}
\end{center}
\end{ruledtabular}
\end{table}

\begin{figure}[h]
\begin{tabular}{cc}
\includegraphics[width=0.50\linewidth]{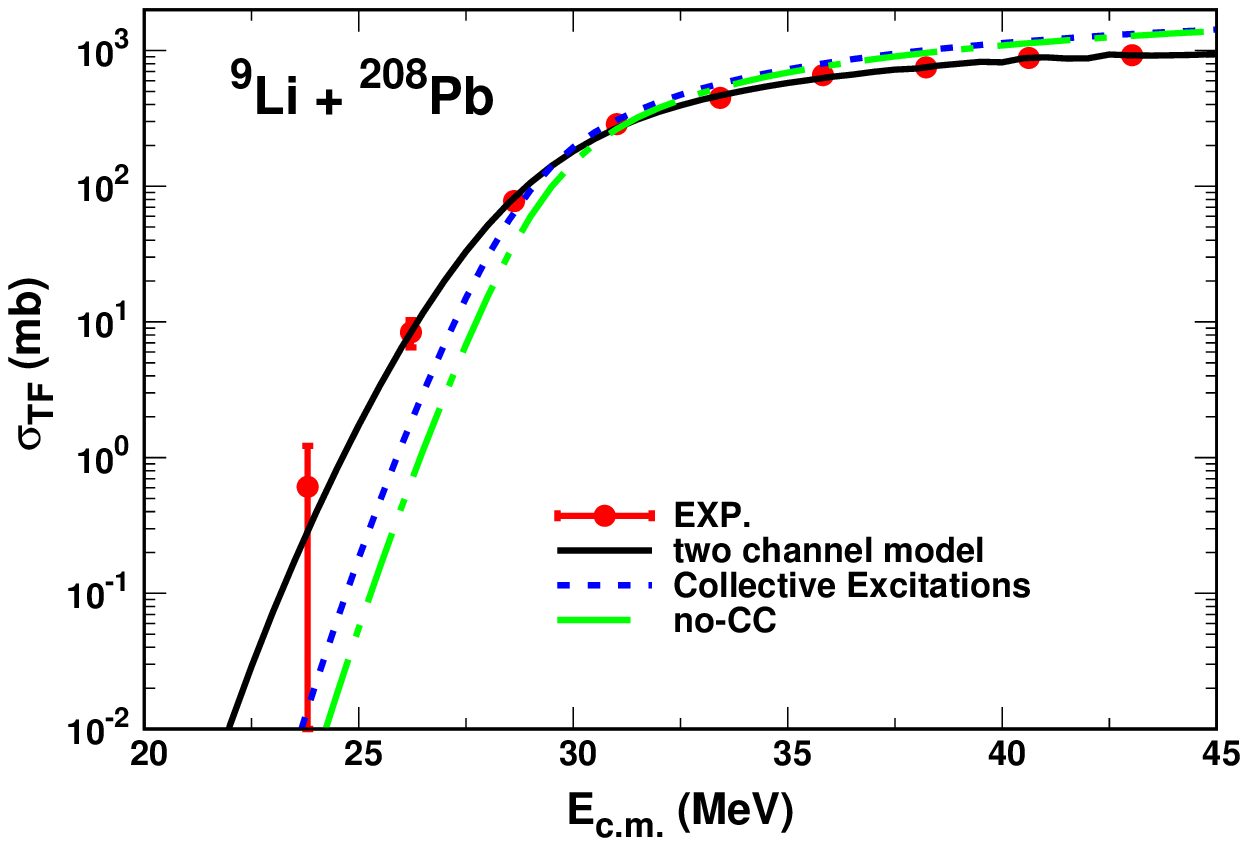} & \includegraphics[width=0.50\linewidth]{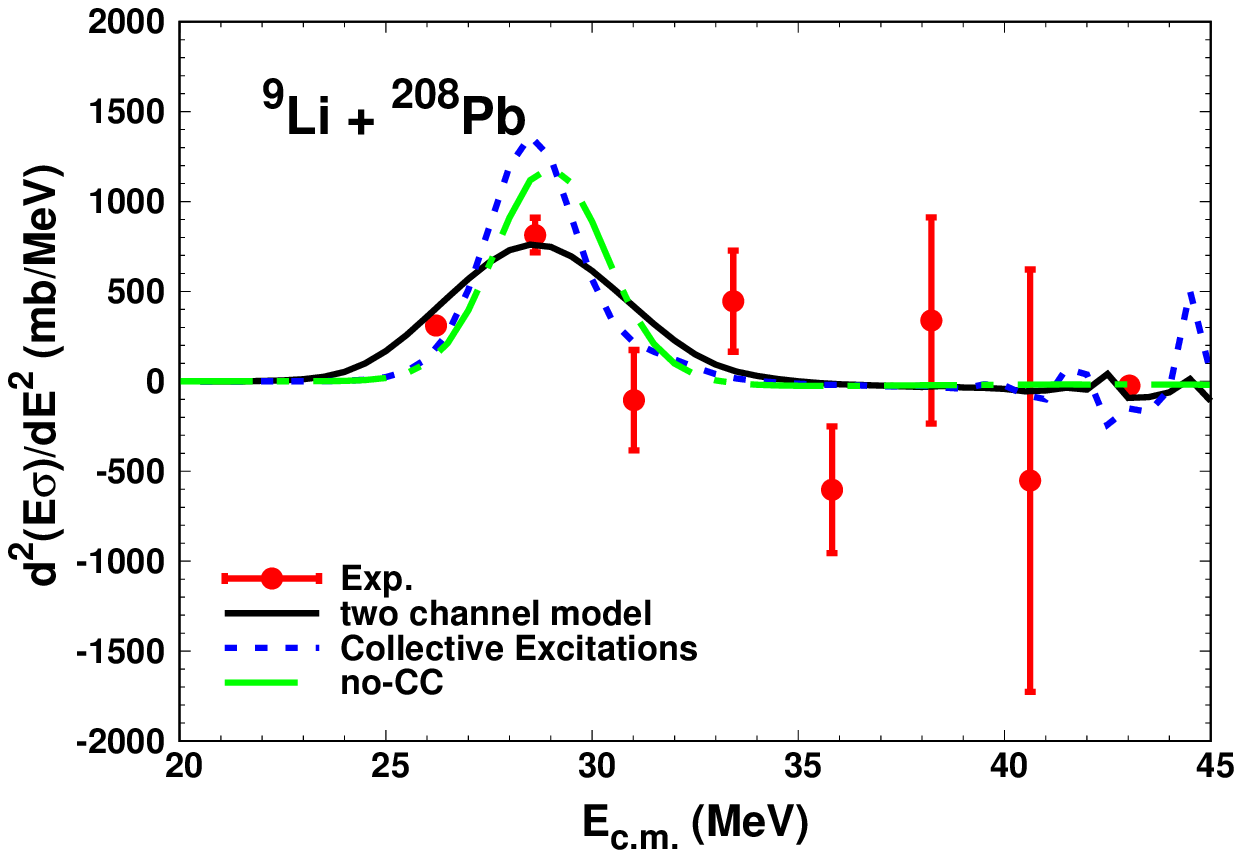} \tabularnewline
\end{tabular}
\caption{
 (Color online)~ The total fusion cross sections (the left panel) and
the fusion barrier distribution, $d^2(E\sigma)/dE^2$ (the right panel)
for the $^{9}$Li + $^{208}$Pb system.
The black solid lines denote the results of the two-channel calculation,
which takes into account the two-neutron transfer process, while
the blue dashed lines are obtained by including the collective excitations
in the colliding nuclei.
The result of the single channel calculation is denoted by
the green dot-dashed lines.
See text for details.
The experimental data are taken from Ref. ~\cite{PhysRevC.80.054609}.}
\label{fig:1}
\end{figure}

The solid (black) line in the left panel of
Fig.~\ref{fig:1} shows the result of the two-channel
calculation for fusion cross sections.
In order to draw this curve, we exploit a fitting process to
the experimental data and find the optimum values for the four adjustable
parameters, that is, the effective $Q$-value, $Q_{23}$, and
$F_t$, $R_{\rm coup}$, and $a_{\rm coup}$ in the coupling form factor,
Eq. (\ref{eq:2}). The optimum value for these parameters
are listed in Table 2.
One can see that the experimental fusion cross sections are well reproduced
with this calculation.
For comparison, the figure also shows the coupled-channels (CC) calculation
with collective excitations in the colliding nuclei (but with no transfer
coupling; the blue dotted line) as well as the single-channel
calculation (the green dot-dashed line).
For the former CC calculation, we include
the rotational excitation to the first excited state at 2.69 MeV in $^9$Li
with the quadrupole deformation parameter of
$\beta_2 = 0.469$~\cite{PhysRevC.87.064320}
as well as the vibrational coupling to the 3$^-_1$ state
in $^{208}$Pb at 2.615 MeV with the deformation
parameter of $\beta_{3}=0.111$.
As has been shown in Ref.~\cite{PhysRevC.80.054609}, the collective
excitations alone do not account well for the experimental data.

In the right panel of Fig.~\ref{fig:1}, we present the fusion
barrier distribution defined by $d^2(E\sigma)/dE^2$
~\cite{ROWLEY199125}.
The experimental fusion barrier distribution is extracted
from the experimental fusion cross sections
using
a point difference formula.
One can see that the barrier distribution is significantly widened
by the transfer coupling. This brings in the low energy strength in
the barrier distribution, which eventually results in the large enhancement
of subbarrier fusion cross sections shown in the left panel of
Fig.~\ref{fig:1}.

\section{$^{11}$L\lowercase{i}+$^{208}$P\lowercase{b} fusion reaction}

We next discuss the fusion of the $^{11}$Li+$^{208}$Pb system.
To this end, we again neglect the collective excitations of the colliding
nuclei and focus on the transfer couplings.
Although the breakup of
$^{11}$Li would also play a role in the fusion reaction, we neglect it in
the present calculation. For weakly bound nuclei, large transfer cross
sections are often observed experimentally at energies below the Coulomb
barrier \cite{Raabe04,Lemasson09},
and thus the role of transfer couplings is expected to be as important
as the breakup channel. Since the role of transfer couplings in the fusion
of weakly bound nuclei has not been well clarified, we here
concentrate on the transfer couplings, leaving the simultaneous treatment
of the transfer and the breakup channels for an interesting future work.

Introducing effective transfer states as in the previous section,
we thus solve the following three-channel problem:
\begin{equation}
\left(\begin{array}{ccc}
K+V_1(r)-E & F_{  1 \rightarrow 2 }(r) & 0 \\
F_{ 1 \rightarrow 2 }(r) & K+V_2(r)-(E+Q_{12})& F_{ 2 \rightarrow 3 }(r) \\
0 & F_{  2 \rightarrow 3 }(r) &K+V_3(r)-(E+Q_{12}+Q_{23}) \\
\end{array}\right)
\left(\begin{array}{c}
\psi_1(r)  \\
\psi_2(r)  \\
\psi_3(r)  \\
\end{array}\right)= ~ 0 ,
\label{eq:1}
\end {equation}
where
the channels $i$= 1, 2 and 3 correspond to the $^{11}$Li+$^{208}$Pb,
$^{9}$Li+$^{210}$Pb and $^{7}$Li+$^{212}$Pb systems, respectively.
In this equation, we have neglected the direct coupling between the channels
1 and 3, as the direct four-neutron transfer process is quite unlikely.

\subsection {Internuclear potential for each channel}

We first determine the inter-nucleus potential for each channel, $V_i(r)$, in
Eq. (\ref{eq:1}).
For the second and the third channels,
we employ the Aky\"uz-Winther potential~\cite{akyuz1981proceedings},
as in the previous section.
The actual values for the parameters are listed in Table 3.
Because the mass number of the projectile and the target
is different only slightly, those values are close to the parameters for
the $^{9}$Li+$^{208}$Pb system listed in Table 1.

\begin{table}[h]
\begin{ruledtabular}
\begin{center}
\caption{Same as Table 1, but for the $^{11}$Li + $^{208}$Pb reaction.
The potential for the channels 2 and 3 is based on the Aky\"uz-Winther
potential, while the parameters for the channel 1 are obtained by fitting the
double folding potential to a Woods-Saxon form. }
\label{tab:2}
\begin{tabular}{c|ccc|ccc}
Channel &   $V_0$ (MeV) & $r_0$ (fm) &   $a_0$ (fm) & $V_b$ (MeV) & $R_b$ (fm) & $\hbar\Omega$ (MeV)  \\ \hline
 $^{11}$Li + $^{208}$Pb ~~(channel 1)      &   90.309 & 1.090  & 0.852 & 27.3 & 12.1
& 3.04             \\
 $^{9}$Li +  $^{210}$Pb ~~(channel 2)     &   47.304 & 1.178  & 0.636 & 29.0 &
11.5 & 4.06            \\
 $^{7}$Li +  $^{212}$Pb ~~(channel 3)     &   47.298 & 1.177  & 0.626 & 29.6 &
11.29 & 4.70            \\
\end{tabular}
\end{center}
\end{ruledtabular}
\end{table}

For the potential for the $^{11}$Li + $^{208}$Pb channel, on the other hand,
one expects a large deviation from the Aky\"uz-Winther potential
due to the halo structure of the
$^{11}$Li nucleus~\cite{choi2017cc}.
In order to take into account the halo structure,
we employ a folding potential approach~\cite{TAKIGAWA199123} and
construct the potential as,
\begin{equation}
V(r) = \int d\textbf{r}_1 \int d\textbf{r}_2 \rho_p(r_1)\rho_t(r_2) v_{NN}
\left(\left| \textbf{r}-\textbf{r}_1+\textbf{r}_2\right|\right),
\label{eq:folding}
\end{equation}
where $\rho_p(r_1)$ and $\rho_t(r_2)$ are the density distribution
for the projectile and target nuclei, respectively.
For the effective nucleon-nucleon interaction, $v_{NN}$,
we here employ
the M3Y interaction~\cite{BERTSCH1977399} with the zero-range approximation
to the knock-on exchange term given by
\begin{equation}
v_{NN}(r)=-2134\,\frac{e^{-2.5r}}{2.5r}+7999\,\frac{e^{-4r}}{4r}-275.81\,
\delta(\textbf{r}),
\end{equation}
where the value of each parameter is given in units of fm or MeV.
For the density distribution for the target nucleus,
we use the Woods-Saxon form,
\begin{equation}
\rho_t(r)=\frac{\rho_0}{1+\exp{\left(\frac{r-c}{z}\right)}},
\end{equation}
with $c$=6.67 fm, $z$=0.545 fm and $\rho_0$= 0.157 fm$^3$
for $^{208}$Pb~\cite{SATCHLER1994241}.
For the projectile nucleus $^{11}$Li,
we assume that
it is comprised of a core
nucleus $^{9}$Li and two valence neutrons
due to the halo structure.
The projectile density is then given by
\begin{equation}
\rho_{p}(r)  = \rho_c(r)+\rho_{2n}(r),
\end{equation}
where $\rho_c(r)$ is the core density and $\rho_{2n}(r)$ is
the density for the valence neutrons.
The folding potential, Eq. (\ref{eq:folding}), is thus also
divided into two
parts,
\begin{equation}
      V(r) = V_{c-t}(r)+V_{2n-t}(r),
\end{equation}
with
\begin{eqnarray}
V_{c-t}(r) &=&\int d\textbf{r}_1 \int d\textbf{r}_2 \,\rho_c(r_1)\rho_t(r_2)
v_{NN}\left(\left| \textbf{r}-\textbf{r}_1+\textbf{r}_2\right|\right), \\
V_{2n-t}(r)&=&\int d\textbf{r}_1 \int d\textbf{r}_{2} \,\rho_{2n}(r_1)\rho_t(r_2)
v_{NN}\left(\left| \textbf{r}-\textbf{r}_1+\textbf{r}_2\right|\right)~.
\\\nonumber
\label{eq:8}
\end{eqnarray}
In the following, we replace
the interaction between the core and the target nuclei, $V_{c-t}(r)$,
with the Aky\"uz-Winther potential
for the $^{9}$Li+$^{208}$Pb system.
For the interaction between the valence neutrons and the target,
following Ref. ~\cite{TAKIGAWA199123}, we employ the dineutron cluster
model and
introduce a Yukawa function for the density of the valence neutrons, that is,
\begin{equation}
\rho_{2n}(s)=\rho_0\,\frac{e^{-2\kappa s}}{s^2}~,
\end{equation}
where $\rho_0=\kappa / \pi$, $\kappa$ being determined from the
two-neutron separation energy, $S_{2n}=0.369$ MeV,
and $s$ is the distance between the core nucleus and the center of mass
of the two valance neutrons.

The solid (black) lines in Fig. 2 show the folding potential for the $^{11}$Li+$^{208}$Pb
so obtained. The left panel shows the valence-target potential, while
the right panel shows the total potential, which includes the
valence-target and the core-target nuclear potentials as well as the Coulomb
potential between the projectile and the target nuclei.

\begin{figure}[h]
\begin{tabular}{cc}
\includegraphics[width=0.50\linewidth]{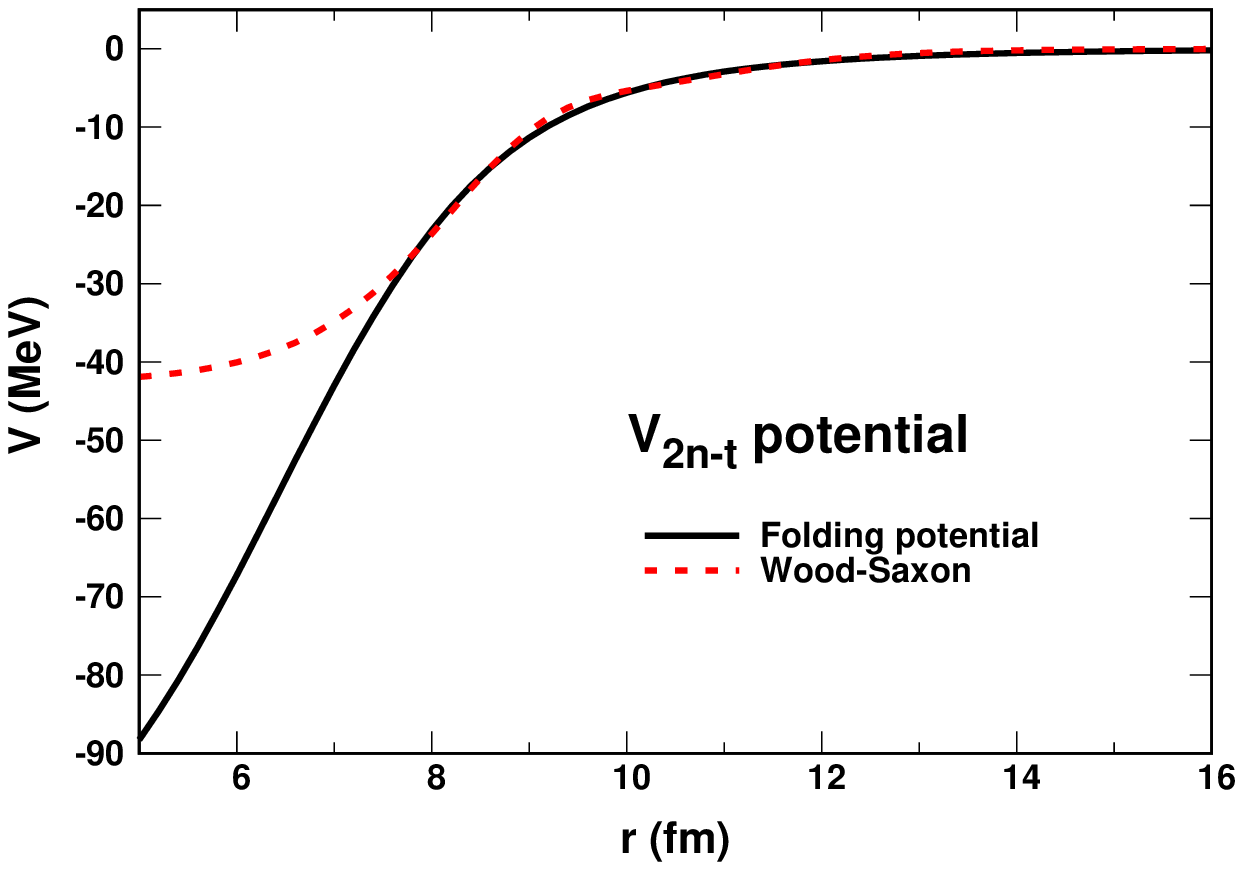} & \includegraphics[width=0.50\linewidth]{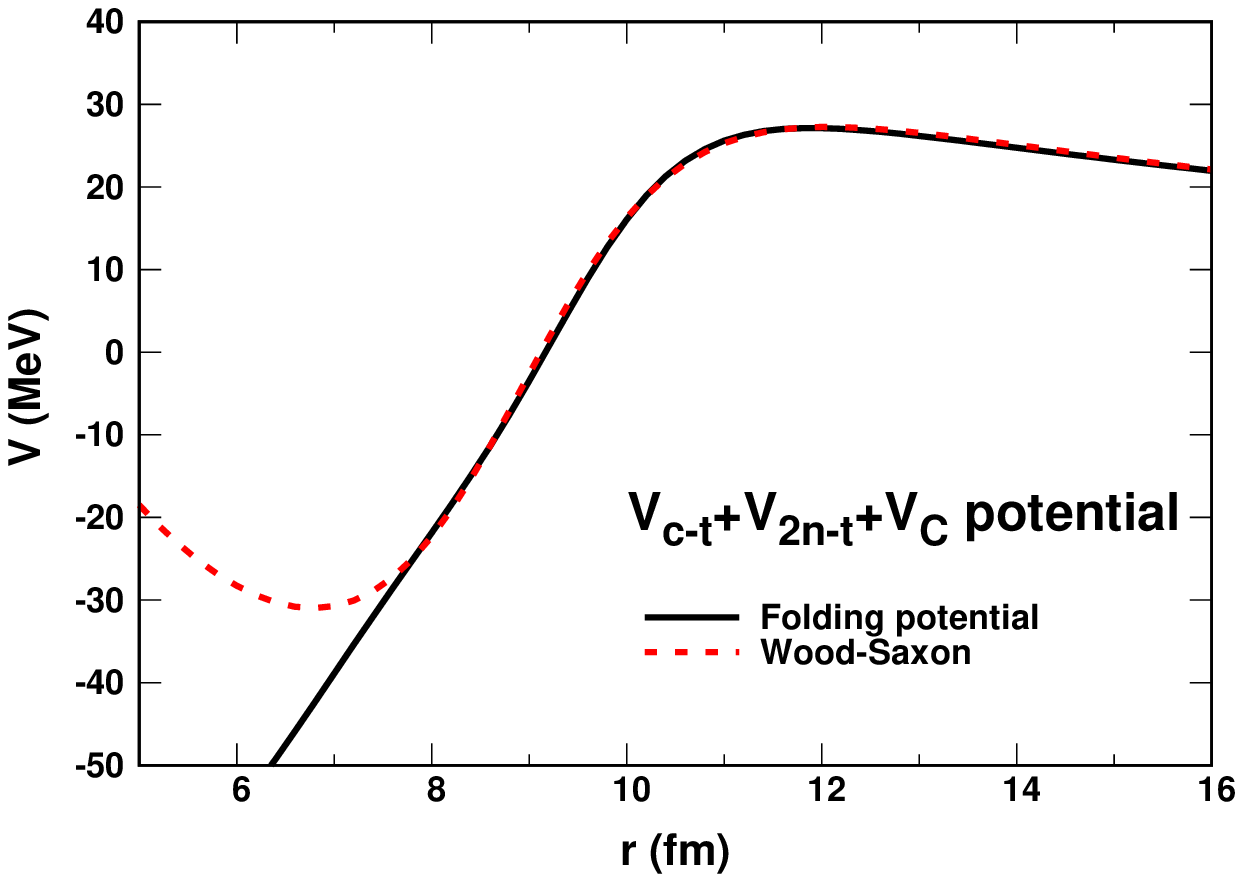} \tabularnewline
\end{tabular}
\caption{(Color online)~The potential for the
$^{11}$Li+ ${}^{208}$Pb fusion reaction. The left panel shows
the potential between the valence neutrons in $^{11}$Li and the target
nucleus, while the right panel is for the total potential including the
Coulomb potential.
The (black) solid lines denote the folding potential, while
the (red) dashed lines show its fit with a Wood-Saxon potential.}
\label{fig:2}
\end{figure}
%

In order to discuss properties of the potential, we fit this potential
with a Woods-Saxon function. The resultant potential is shown in Fig. 2
by the dashed (red) lines, whose parameters are listed in Table 3.
One can see that the folding potential can be well fitted with the
Woods-Saxon potential for $r$ larger than 8 fm. Note that fusion
cross sections are
insensitive to the details of the nuclear potential for the smaller values
of $r$, due to the strong absorption inside the barrier, and thus we
actually use
the fitted Woods-Saxon potential in the calculations presented in the next
subsection.

Because of the halo structure of $^{11}$Li, the potential for the
$^{11}$Li+$^{208}$Pb system shows different behavior compared to the
potential for the other systems. Since the Coulomb barrier is lowered,
the potential is deeper than those potentials for the channels 2 and 3.
Moreover, the value of diffuseness parameter, $a_0$, is significantly
larger, reflecting the extended density distribution of the $^{11}$Li nucleus.
This results in a smaller value of the barrier curvature, as shown in
Table 4.
We mention that this feature
has been treated as a long-range interaction
in Ref.~\cite{PhysRevC.90.054615}.

\subsection {Results of the coupled-channels calculations}

\begin{figure}[h]
\begin{tabular}{cc}
\includegraphics[width=0.50\linewidth]{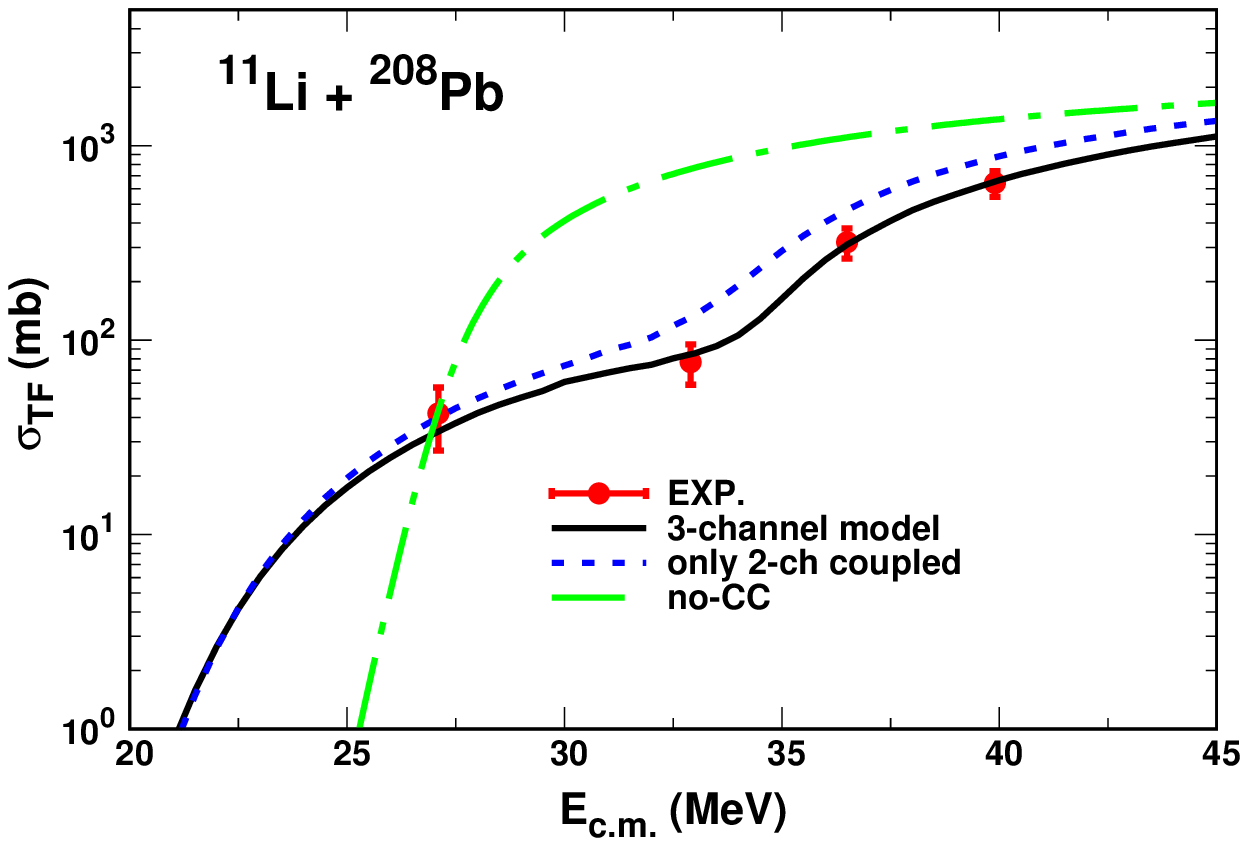} & \includegraphics[width=0.50\linewidth]{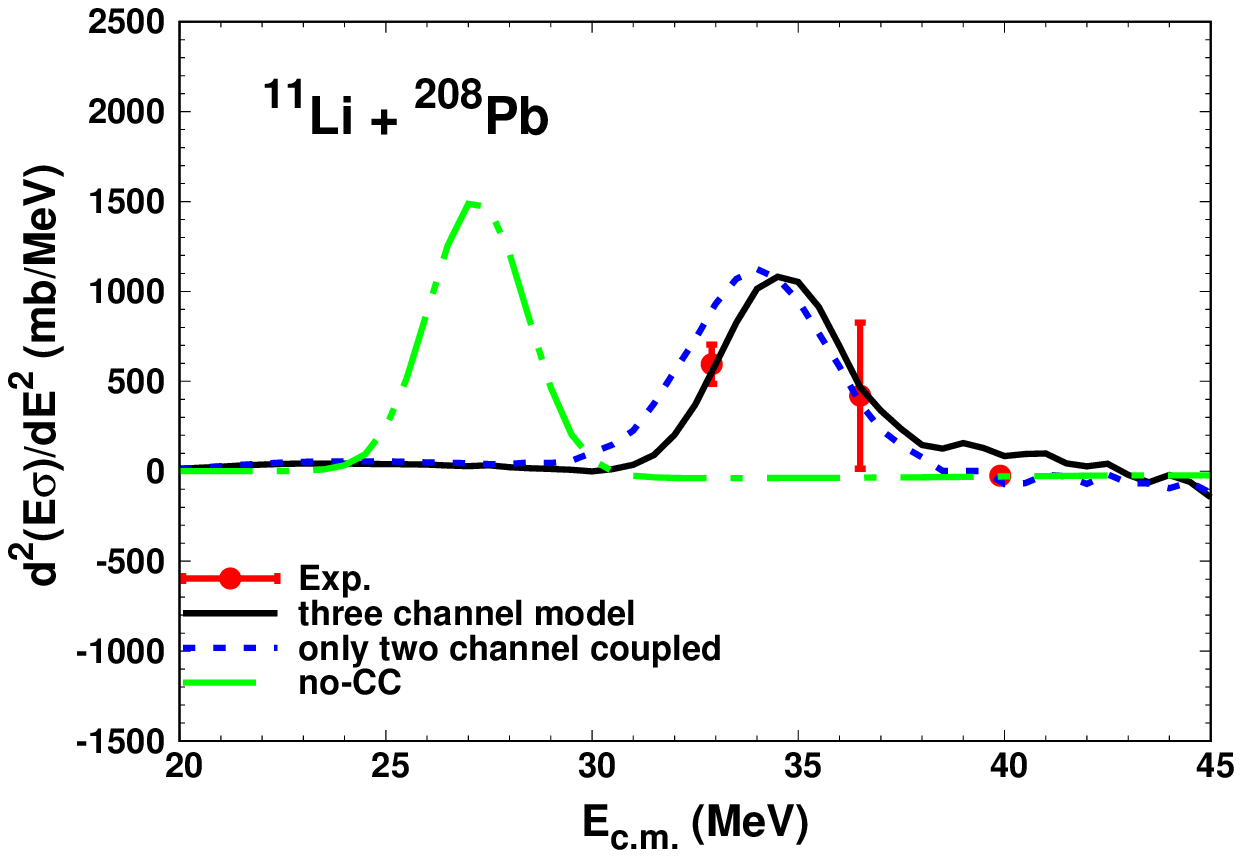} \tabularnewline
\end{tabular}
\caption{(Color online)~
The total fusion cross sections (the left panel) and the fusion
barrier distribution, $d^2(E\sigma)/dE^2$, (the right panel) for
the $^{11}$Li+ ${}^{208}$Pb system.  The (black) solid lines show the
results of the three-channel coupling model with the
$^{11}$Li+ ${}^{208}$Pb,
$^{9}$Li+ ${}^{210}$Pb,
and $^{7}$Li+ ${}^{212}$Pb channels,
while the (blue) dotted lines show the result
obtained by switching off the transfer coupling between
$^{9}$Li+ ${}^{210}$Pb
and $^{7}$Li+ ${}^{212}$Pb.
The (green) dot-dashed lines show the results of the single-channel
calculation.
The experimental data are taken from Ref.~\cite{PhysRevC.87.044603}.}
\label{fig:3}
\end{figure}

Finally, we solve the coupled-channels equations, Eq. (\ref{eq:1}),
and calculate fusion cross sections for the $^{11}$Li+$^{208}$Pb system.
In order to reduce the number of adjustable parameters, we assume
that the effective $Q$-value, $Q_{23}$, and the parameters for the
transfer coupling form factor for the coupling between the channels 2 and 3
are the same as those determined in the previous section (see Table 2),
even though the mass numbers are slightly different.
The number of adjustable parameter is now
reduced to four,
that is, the effective $Q$-value and the parameters for the form factor
for the coupling between the channels 1 and 2.
These are determined by fitting the calculated result for fusion
cross sections to the experimental data, as has been done in the
previous section for the $^{9}$Li+$^{208}$Pb system.
The resultant fusion cross sections are shown in the (black) solid
line in the left panel of Fig.~\ref{fig:3}, whereas
the resultant parameters are shown in Table 4.
The theoretical uncertainties for $Q$ and $F_t$ for the coupling from the channel 1 to 2 are estimated to be $Q=+8.346^{+1.163}_{-0.897}$ MeV and $F_t=40.227^{+2.45}_{-2.136}$ fm, respectively, for which we have used the $\chi$-square fitting with the fixed values for $r_{coup}$ and $a_{coup}$.


%
\begin{table}[h]
\begin{ruledtabular}
\begin{center}
\caption{
Same as Table 2, but for the $^{11}$Li + $^{208}$Pb reaction.
The parameters for the coupling between the channels 1 and 2 are obtained
by fitting to the experimental data, while those for the coupling between
the channels 2 and 3 are taken to be the same as those in Table 2.
The $Q$-value
for the ground-state-to-ground state transfer between the channels 1 and 2
is $Q_{gg}$= +8.852 MeV.
}
\begin{tabular}{c|cccc}
Channel               & $Q$ (MeV) & $F_{t}$ (MeV fm) & $r_{\rm coup}$ (fm)
&   $a_{\rm coup}$ (fm)   \\ \hline
 $1 \rightarrow 2 $  & +8.346  & 40.227   & 1.666  & 0.857              \\
 $2 \rightarrow 3 $  & $-$3.204 & 51.367   & 1.357 & 0.264               \\
\end{tabular}
\end{center}
\end{ruledtabular}
\end{table}

For comparison, the figure also shows the results of the single-channel
calculation (the (green) dot-dashed line).
One can see that the fusion cross sections are largely overestimated
in this calculation at energies above the barrier.
But, if we take the neutron-transfer channel into account,
they becomes compatible with the experimental data, as shown in the dashed (blue) line. This is due to the transfer coupling with a large positive
$Q$-value \cite{vonOertzen96,Zagr03},
for which the higher energy peak in the barrier distribution carries
more weight than the lower energy peak, as is evident in the
right panel in Fig. 3. The Coulomb barrier is thus effectively shifted
towards high energy, reducing the fusion cross sections at energies
above the barrier.
At energies below the Coulomb barrier, on the other hand,
fusion cross sections in the coupled-channels calculations
are largely enhanced as compared to the fusion cross sections in
the single-channel calculation due to
the lower
energy peak in the barrier distribution, even though it carries
only a small weight.

In order to investigate the role of the third channel, we also present by the (blue) dotted lines the result of the two-channel calculation obtained by switching
off the coupling between the channels 2 and 3.
One can see that the coupling to the
$^{7}$Li+$^{212}$Pb channel, that is, the multi two-neutron transfer channel,
plays a significant role, even though the main effect comes from the
single two-neutron transfer channel, {\it i.e.,} the coupling between the
channels 1 and 2.

\section{Conclusion}

We have calculated total fusion cross sections for the
$^{9,11}$Li +$^{\text{208}}$Pb systems, for which
the $^{11}$Li nucleus has a halo structure,
by taking into account multiple two-neutron transfers in the
coupled-channels approach.
To this end, we have constructed the nuclear potential for the
$^{11}$Li +$^{\text{208}}$Pb channel with the double folding procedure based
on the dineutron cluster model.
We have employed the global Ak\"uz-Winther potential for
all the other channels.
By adjusting the effective transfer $Q$-values and the parameters for the
coupling form factors, we have successfully reproduced the experimental
fusion cross sections for both the systems simultaneously.
This clearly indicates that (multi-) neutron transfer channels owing to the positive $Q$ value, specifically for $^{11}$Li channel, play
an important role in fusion of weakly bound nuclei. We did not include the breakup channel in this work since we expected that the breakup channel does not affect much cross-sections for total fusion at energies above the Coulomb barrier owing to the large gap between the Coulomb barrier and the breakup channel $Q$ value.

For fusion of the $^{11}$Li +$^{\text{208}}$Pb system,
the experimental data exist only at four energy points.
This has prevented us to uniquely determine the parameters, especially the
effective $Q$-values for the transfer couplings.
In order to gain a deeper insight into the role of transfer couplings in
fusion of weakly bound nuclei,
it would be helpful if total fusion cross sections
for this system will be measured
in near future at more data points, especially at energies below the
Coulomb barrier.

\section*{ACKNOWLEDGMENT}
This work was supported by the National Research Foundation of Korea (Grant Nos. NRF-2016R1C1B1012874, NRF-2014R1A1A2A16052632, NRF-2015R1D1A3A01017378, NRF-2015R1A2A2A01004727, NRF-2015K2A9A1A06046598, and NRF-2017R1E1A1A01074023).
K.H. thanks Soongsil University and Kangwon National University at Dogye
for their hospitality and for financial supports for his visit to
those universities.


\end{document}